\documentclass[prl,aps,twocolumn,showpacs,floatfix]{revtex4-1}
\usepackage{amsfonts}
\usepackage{amssymb}
\usepackage{hyperref}
\usepackage{graphicx}
\usepackage{dcolumn}
\usepackage{bm,amsmath,verbatim}
\usepackage{mathrsfs}
\usepackage{color}

\usepackage[T1]{fontenc}
\usepackage[latin9]{inputenc}
\usepackage{booktabs}

\hypersetup{colorlinks,
linkcolor=blue,          citecolor=blue,        filecolor=blue,      urlcolor=blue           }

\newcommand{\ket}[1]{|#1\rangle}
\newcommand{\bra}[1]{\langle #1 |}

\newcolumntype{x}[1]{>{\centering\let\newline\\\arraybackslash\hspace{0pt}}p{#1}}
\newcolumntype{I}{!{\vrule width 1pt}}

\begin{document}

\title{Symmetry-Protected Topological Phases in a Rydberg Glass}
\author{Kai Li$^{1}$}
\author{Jiong-Hao Wang$^{1}$}
\author{Yan-Bin Yang$^{1}$}
\author{Yong Xu$^{1,2}$}
\email{yongxuphy@tsinghua.edu.cn}
\affiliation{$^{1}$Center for Quantum Information, IIIS, Tsinghua University, Beijing 100084, People's Republic of China}
\affiliation{$^{2}$Shanghai Qi Zhi Institute, Shanghai 200030, People's Republic of China}

\begin{abstract}
Recent theoretical studies predict that structural disorder, serving as a bridge connecting a crystalline material to an amorphous material,
can induce a topological insulator from a trivial phase. However, to experimentally observe such a topological
phase transition is very challenging due to the difficulty in controlling structural disorder in a quantum material. 
Given experimental realization of randomly positioned Rydberg atoms, such a system is naturally suited to 
studying structural disorder induced topological phase transitions and topological amorphous phases.
Motivated by the development,
we study topological phases in an experimentally accessible one-dimensional amorphous Rydberg atom 
chain with random atom configurations. In the single-particle level, we find symmetry-protected topological amorphous 
insulators and a structural disorder induced topological phase transition, indicating that Rydberg atoms provide an ideal platform
to experimentally observe the phenomenon using state-of-the-art technologies. Furthermore, we predict the existence of a gapless symmetry-protected topological phase of interacting bosons
in the experimentally accessible system.
The resultant many-body topological amorphous phase is characterized by a $\mathbb{Z}_2$ invariant. 
\end{abstract}
\maketitle

Although topological phases of matter are primarily pursued in crystalline
materials with translational symmetry~\cite{Kane2010RMP,Zhang2011RMP,Chiu2016RMP,VishwanathRMP,XuReview,ZhuReview}, 
recent studies showed that topological phases
for non-interacting quantum particles can also exist in two or three dimensional amorphous
systems~\cite{Agarwala2017PRL,Mansha2017PRB,YB2019PRL,Ojanen2018NC,Irvine2018NP,Hellman2019arxiv,
	Agarwala2020PRR,Prodan2018JPA,Zhang2019PRB,Chern2019EPL,
	Fazzio2019NL,Bhattacharjee2020PRB,Sahlberg2020PRR,
	Marsal2020arxiv,Liu2020Research,Ojanen2020PRR,Zhang2020LSA,Grushin2020arxiv,Spring2020arXiv,Lewenkopf2020arxiv}.
Such systems have randomly distributed lattice sites corresponding to 
a limiting case with maximum structural disorder arising from atom position randomness. 
Remarkably, it has been theoretically shown that such disorder can induce a topological phase 
transition in three dimensions~\cite{Jionghao2021,Griffin2020arxiv}, reminiscent of topological Anderson insulators~\cite{Shen2009PRL}, 
a topological phase induced by onsite disorder.
However, 
it is very challenging to experimentally observe such a structural disorder induced topological phase transition in a 
quantum material.

Besides fermionic systems, bosonic systems can support
symmetry-protected topological (SPT) phases when strong interactions between particles are considered.
In fact, based on topological properties of a quantum many-body ground state, substantial progress has been 
made toward classifying interacting bosonic SPT phases for
gapped systems~\cite{Chen2011PRB,Cirac2011PRB,Chen2012Science,Chen2013PRB,Levin2012PRB,Yuanming2012PRB,Levin2013PRL}. 
Although the classification does not necessarily require the existence of translational symmetry,
it is not clear whether SPT phases for interacting bosons can exist
in amorphous systems.

Rydberg atoms have proven to be a powerful platform for quantum simulation and quantum computation due to their high controllability and huge dipolar interactions~\cite{2010RMP_Rydberg,Schauss2018QST,2020NP_Rydberg}. 
A variety of quantum spin models and topological models can be simulated in a Rydberg atom platform~\cite{Hernandez2005PRA, Lesanovsky2012PRL, Dauphin2012PRA, Xiaopeng2015NC, Norman2017PRL, Brune2018PRX, Papic2018NP, Papic2018PRB, Buchler2018QST, Sachdev2020PRL, Dalmonte2020PRX, Zoller2020PRX, Khazali2021}, and
several of these models have been experimentally realized~\cite{Adams2015PRL, Browaeys2016Nature, Lukin2017Nature, Bakr2018PRX, Lienhard2018PRX, Browaeys2019Science}.
In particular, a bosonic version of the Su-Schrieffer-Heeger (SSH) model has recently been experimentally engineered with Rydberg atoms, leading to an observation of SPT phases of interacting bosons in regular lattices~\cite{Browaeys2019Science}. 
Meanwhile,
the development of experimental techniques enables experimentalists to trap Rydberg atoms individually in any position in space using optical tweezers~\cite{Browaeys2016Science,Lukin2016Science}.
In fact, structural disorder has been realized in experiments by trapping a cloud of randomly positioned atoms in an optical trap~\cite{Orioli2018PRL,Signoles2021PRX}.
Such development makes Rydberg atoms a natural platform to study topological phases in amorphous lattices and structural disorder induced topological phase 
transitions.   

Motivated by the development, we study the SPT phases in a one-dimensional (1D) amorphous bosonic model with long-range hopping based on the experimental setup.
In the single-particle level, we show that the topological phase can exist in amorphous lattices; the topological properties are
characterized by the polarization, the boundary charge and the local density of states (LDOS).
Remarkably, we also find the structural disorder induced topological phase transitions in the system. 
In the many-body level with hard-core bosons at half-filling, the topological property of a ground state is characterized by a $\mathbb{Z}_2$ index, which is protected to be quantized by time-reversal symmetry, particle-hole symmetry or another anti-unitary symmetry. 
Through numerically calculating the ground state by exact diagonalization (ED) and matrix product
state (MPS)~\cite{Orus2014AP,Reiher2015JCP}, we show that, in contrast to the single-particle case,
the ground state of interacting bosons exhibits a large intermediate regime with the coexistence of topologically trivial and nontrivial states. 
Yet, further delicate finite-size analysis suggests the existence of topological amorphous phases in the many-body case. 
In both the single-particle and many-particle cases, we demonstrate how to experimentally observe the topological phases using 
a global microwave pulse in a realistic Rydberg platform.   

\begin{figure}[t]
	\includegraphics[width=3.3in]{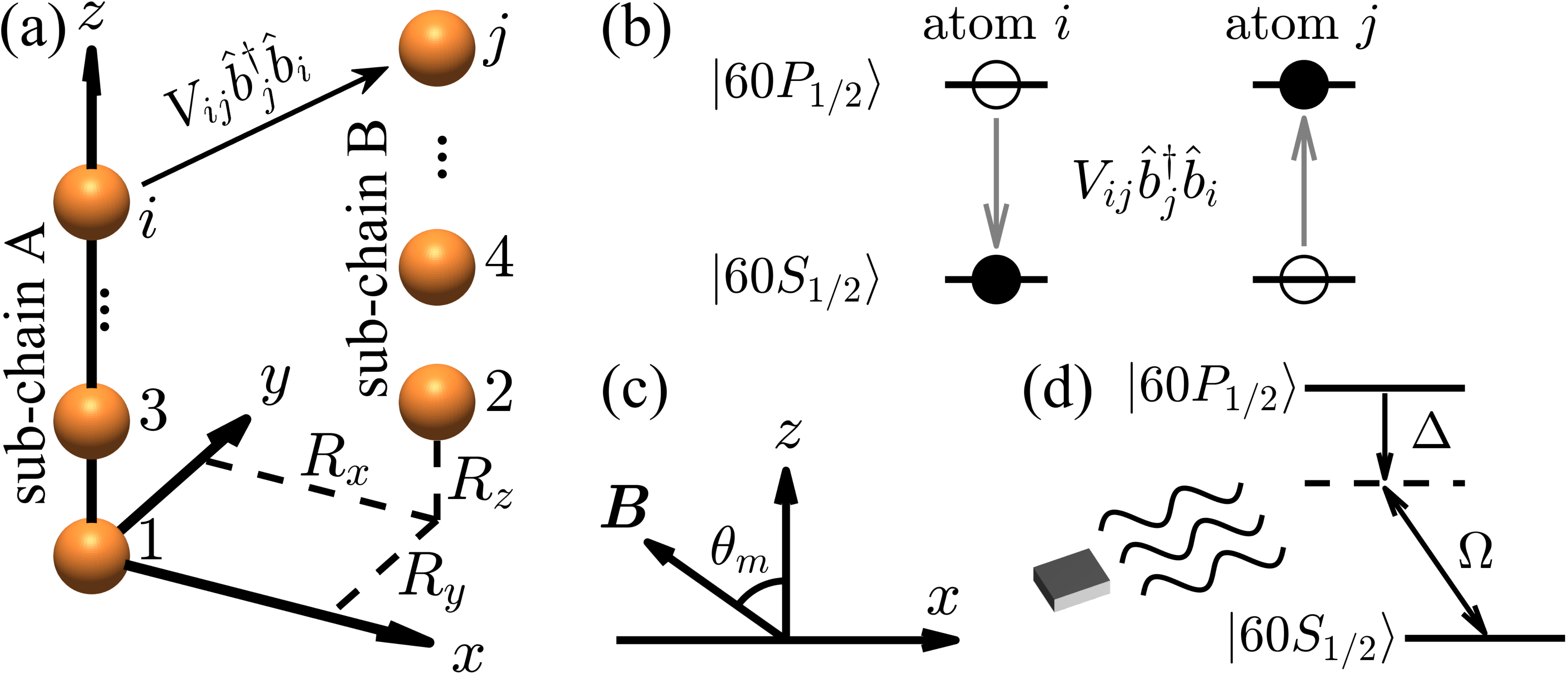}
	\caption{(Color online)
		(a) Schematics of our system consisting of two sub-chains of Rydberg atoms with atoms $2i-1$ and
		$2i$ forming a unit cell.
		Here ${\bm{R}}=(R_x,R_y,R_z)$ is the position vector between two atoms in a unit cell.
		(b) Dipolar interactions between two atoms lead to the hopping of an excitation between these two atoms. 
		For example, $V_{ij}\hat{b}_j^\dagger \hat{b}_i$ indicates the hopping of an excitation from atom~$i$ to atom~$j$.
		(c) The magnetic field $\bm{B}$ lies in the $(x,z)$ plane with the polar angle $\theta_m=\arccos (1/\sqrt{3})$ such that the hopping between atoms in a sub-chain vanishes.
		(d) Microwave fields coupling the two Rydberg states with the Rabi frequency $\Omega$ and detuning $\Delta$ for experimental observations of the topological phases.
		}
	\label{fig1}
\end{figure}

\emph{Model Hamiltonian.}---
We start by considering a chain of Rydberg atoms comprised of two sub-chains with $2N$ atoms as shown in Fig.~\ref{fig1}(a).
For each atom, we consider two Rydberg states: an $s$-level (e.g., $|60S_{1/2},m_J=1/2\rangle$) and a $p$-level 
(e.g., $|60P_{1/2},m_J=-1/2\rangle$). 
Because of the dipolar interaction between two atoms that couples these states, we can use the following 
Hamiltonian to describe the system, 
\begin{equation}
	\hat{H} = \sum_{i<j}^{2N} V_{ij} (\hat{b}_i^{\dagger} \hat{b}_j + \hat{b}_j^{\dagger} \hat{b}_i),
	\label{Hmodel}
\end{equation}
where $\hat{b}_i^\dagger$ ($\hat{b}_i$) creates (annihilates) a hard-core boson at site $i$ [see Fig.~\ref{fig1}(b)],
and $\hat{b}_i^\dagger \ket{0}$ ($\ket{0}$ is the vacuum state where all atoms are in the s-level) denotes the state where only the $i$-th atom is excited to the p-level.
Since an atom can only be excited to the p-level once, it naturally realizes a hard-core boson with $(\hat{b}_j^\dagger)^2=0$. 
The hopping amplitude due to the dipolar interaction is $V_{ij}=d^2(1-3 \cos^2 \theta_{ij})/R_{ij}^3$ which depends on the dipole moment $d$ of the Rydberg atom and the angle $\theta_{ij}$ between the magnetic field ${\bm{B}}$ and the position vector ${\bm{R}}_{ij}$ from site $i$ to $j$. 
We note that such a Hamiltonian has been experimentally realized with $^{87}$Rb atoms~\cite{Browaeys2019Science}.

We study the topological properties in a Rydberg glass by randomly placing $N$ unit cells in a 1D box of size $N$;
each unit cell contains an atom in a sub-chain $A$ (labeled by odd numbers) and an atom in a sub-chain $B$ (labeled by even numbers) separated by a vector ${\bm R}=(R_x,R_y,R_z)$ [see Fig.~\ref{fig1}(a)]. 
The hopping within all unit cells is given by $J^\prime=V_{2i-1,2i}=d^2(R_y^2-R_x^2+2\sqrt{2} R_x R_z)/R^5$ with $i=1,2,\cdots,N$. 
Thanks to the angular dependence for the dipolar interaction, we can realize chiral (sub-lattice) symmetry by arranging the atoms aligned along a direction so that its angle with respect to the magnetic field is equal to the `magic angle', i.e., $\theta_m=\arccos (1/\sqrt{3})$, leading to vanishing hopping along each sub-chain.

To investigate the effects of structural disorder on the topological property, we randomly displace atoms from their original regular positions according to
 ${z_{2i-1}}\rightarrow i-1+\delta z_i$ and ${z_{2i}}\rightarrow i-1+R_z+\delta z_i$ with $\delta z_i$ uniformly sampled in the interval $[-W/2,W/2]$. 
When $W=N$, the system becomes completely random.  

In a Rydberg atom experiment, either one particle or many particles~\cite{Browaeys2019Science} can be excited by applying a global microwave field that couples the two 
Rydberg states [see Fig.~\ref{fig1}(d)].  
Motivated by this flexibility, we will study the topological properties in both single-particle and many-particle cases in a Rydberg glass. 
In the following, we set $a_0=1$ and $d^2/a_0^3=1$ as the units of length and energy, respectively.
Because of the disorder feature, all the quantities are averaged over $200$ or more random configurations in numerical calculations. Configuration averaged quantities are denoted by $\overline{\cdots}$.

\emph{Single-Particle Case.}--- 
We now study the scenario with only one excitation in the Rydberg atom chain. 
In this case, the system is described by a single-particle Hamiltonian $H^\mathrm{S}$ with $[H^\mathrm{S}]_{ij}=V_{ij}(1-\delta_{ij})$ ($1 \le i,j \le 2N$)
under a basis $\beta=\{\hat{b}_1^\dagger|0\rangle,\hat{b}_2^\dagger|0\rangle,\cdots,\hat{b}_{2N}^\dagger|0\rangle\}$.  
Since there are no intrachain hopping at the magic angle in the system, $H^\mathrm{S}$ preserves chiral symmetry, i.e., $\Pi H^\mathrm{S} \Pi^{-1}=-H^\mathrm{S}$ with the chiral symmetry operator $\Pi = \mathrm{diag} \{(-1)^{j-1} \}_{j=1}^{2N}$.
$H^\mathrm{S}$ thus belongs to the $\mathbb{Z}$ classification, and its topological property manifests in the existence of zero-energy edge states~\cite{Chiu2016RMP}.

To characterize the topology of the single-excitation Rydberg chain, we calculate the polarization~\cite{Resta1998PRL}
\begin{equation}
	P_\mathrm{S} = [\frac{1}{2\pi} \mathrm{Im} \ln \det (U^\dagger D U)
	      -  \frac{1}{2N} \sum_{i =1}^{2N} x_i ]\mod 1,
	\label{polar_f}
\end{equation}
where $U=(\ket{u_1},\ket{u_2},\cdots, \ket{u_N})$ with $\ket{u_j}$ ($1 \le j \le N$) being eigenstates of the single-particle Hamiltonian $H^\mathrm{S}$ with negative energies under periodic boundary conditions, and $D = \mathrm{diag} \{e^{2\pi i x_j/N} \}_{j=1}^{2N}$ with ${x}_{i}$ being the position of atom $i$.
With chiral symmetry, the polarization $P_\mathrm{S}$ is quantized to zero or $0.5$~\cite{Suppl} and hence can be used as a topological invariant to characterize the topological property of our amorphous system. 

\begin{figure}[t]
	\includegraphics[width=3.3in]{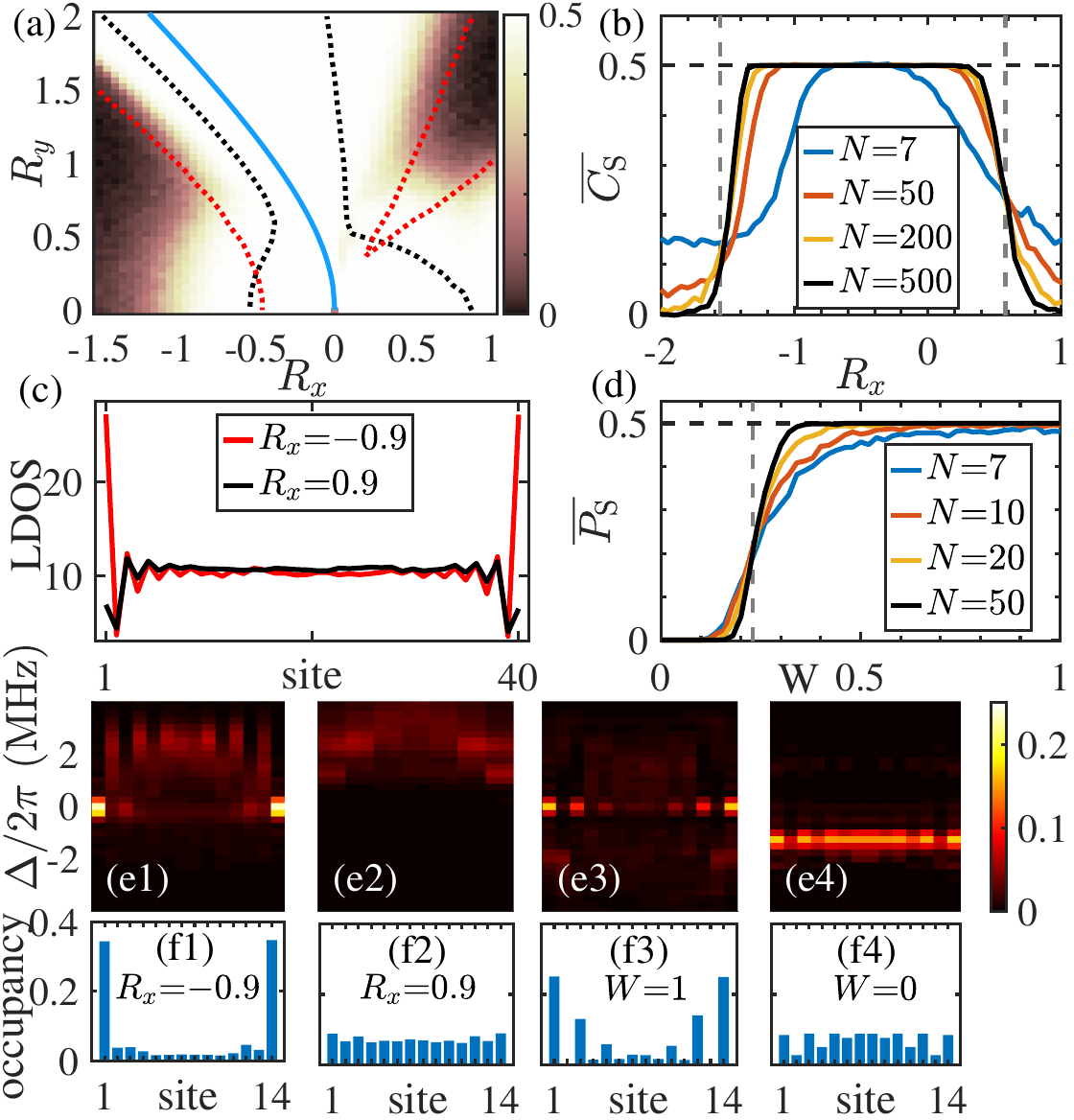}
	\caption{(Color online) 
		(a) Polarization $\overline{P_\mathrm{S}}$ versus $R_x$ and $R_y$ for an amorphous lattice with size $N=200$. 
		The black and red dotted lines show the phase boundaries of an amorphous system with only NN hopping and a regular system with long-range hopping, respectively. 
		The cyan line highlights the parameter region where the intracell hopping vanishes ($J'=0$). 
		(b) Boundary charges $\overline{C_\mathrm{S}}$ versus $R_x$ for distinct system sizes $N$ when $R_y=1.6$.
		(c) Zero energy LDOS for topologically nontrivial ($R_x=-0.9$) and trivial ($R_x=0.9$) phases when $R_y=1.6$ and $N=20$.
		(d) Polarization $\overline{P_\mathrm{S}}$ versus the disorder strength $W$ when $R_x=-0.5$ and $R_y=0$. 
			(e) Occupancy of each site with respect to the microwave detuning $\Delta$.
			(f) Post-selection occupancy distribution when $\Delta=0$.
			(e1,f1) and (e3,f3) [(e2,f2) and (e4,f4)] correspond to the topological (trivial) phase with $R_x=-0.9$ ($R_x=0.9$) in (b) and $W=1$ ($W=0$) in (d), respectively.
			In (e-f), the system size $N=7$. 
			In (a-f), $R_z=0.8$. 
	}
	\label{fig2}
\end{figure}

In Fig.~\ref{fig2}(a), we map out the phase diagram with respect to $R_x$ and $R_y$ based on the polarization.
Clearly, we see a large regime with $\overline{P_\mathrm{S}} \approx 0.5$, showing the existence of topological amorphous phases in a Rydberg glass.
While the system is gapless in both trivial and nontrivial phases due to the strong structural disorder in an amorphous system, all states are localized~\cite{Suppl}.

To understand why topological phases can arise in a Rydberg glass, we consider a simpler model with only nearest-neighbor (NN) hopping and ask whether such a system can host a topological phase in an amorphous geometry. 
The simpler model allows us to analyze a limiting case where the hopping within a unit cell vanishes ($J'=0$).
Evidently, the first and the last sites are isolated without coupling to other sites, giving rise to two zero-energy edge modes. 
These modes occur even when unit cells are randomly distributed.
For the real Rydberg system, since the interchain hopping amplitude decays algebraically with respect to the separation as $1/R_{ij}^3$, we expect that the NN hopping still dominates, and topological phases can also appear in a random Rydberg chain. 
In fact, we find that the long-range hopping significantly enlarges the regime of topological phases as shown in Fig.~\ref{fig2}(a), where the phase boundaries for a Hamiltonian with only NN hopping are also plotted.

To diagnose the topological property of the system, we also compute the boundary charge defined as $C_\mathrm{S} = \sum_{i=1}^{N} (\rho_i - {1}/{2})$ with $\rho_i=\sum_{j=1}^{N} |[\ket{u_j}]_i|^2$ being the local charge density at site $i$ for all states $\ket{u_j}$ with negative energies. 
In the calculation, we add very small onsite potential $\mp \delta$ to the initial and end sites in Hamiltonian $H^\mathrm{S}$, respectively, to lift the degeneracy of edge states.
In Fig.~\ref{fig2}(b), we plot the boundary charge as a function of $R_x$ with $R_y=1.6$
for different system sizes, showing a sharp increase of $\overline{C_\mathrm{S}}$ from zero to $0.5$ near $R_x=-1.56$ and $R_x=0.58$ in
agreement with the phase diagram in Fig.~\ref{fig2}(a).
To further identify that the observed amorphous phase is topological, we display the zero-energy LDOS for states in the topological regime in Fig.~\ref{fig2}(c),
exhibiting large values at two edges, in contrast to small values for states in the trivial regime. It indicates that edge states arise in the topological regime.

In Fig.~\ref{fig2}(a), we also observe that there exist some parameter regions where an amorphous system is in a 
topologically nontrivial phase while a regular system is in a trivial phase, e.g., when $R_x=-0.5$ and $R_y=0$.
It implies that structural disorder can drive a topological phase transition. 
Indeed, we remarkably find that as the disorder strength $W$ increases, the system changes from a topologically trivial phase to a nontrivial one around $W\approx0.23$, as shown in Fig.~\ref{fig2}(d) (see the Supplementary Material for other types of structural disorder~\cite{Suppl}). 

We now show how to experimentally identify the topological phases.
Similar to experimental measurements in the regular case~\cite{Browaeys2019Science}, a weak global microwave field is applied 
to couple the two Rydberg levels for a period of time, 
which can create an excitation when the microwave detuning $\Delta$ matches the energy of the excitation~\cite{Suppl}. At the end, we measure the atom occupancy distribution 
on the $p$-level. Our numerical results demonstrate that in the topological phases, the sites occupancy of the final state exhibits bright peaks 
at the boundaries at the zero detuning [see Fig.~\ref{fig2}(e1) and (e3)], revealing the existence of zero-energy edge modes. Such localized peaks do not appear at the zero detuning 
in the trivial phases [see Fig.~\ref{fig2}(e2) and (e4)]. Figure~\ref{fig2}(f1-f4) further displays the occupancy distribution at zero detuning by postselecting the results corresponding to a single excitation. It illustrates that the excitation in the topological phases mainly resides at edges, whereas in the trivial phases, it is approximately uniformly distributed over all sites.

\emph{Many-Body Case.}--- 
Next, we study the topological property of Hamiltonian (\ref{Hmodel}) in the many-body level.
This Hamiltonian can also be written as an XY spin model with long-range coupling, $\hat{H}=\sum_{i<j}^{2N} V_{ij}(\sigma_i^{+}\sigma_j^{-}+\sigma_j^{+}\sigma_i^{-})$
with $\sigma_j^{\pm}=(\sigma_j^x \pm i\sigma_j^y)/2$ and $\sigma_j^{s}$ ($s=x,y,z$) being the Pauli
matrices at site $j$. The ground state of this spin model corresponds to the ground state of the hard-core bosonic Hamiltonian at half-filling.

\begin{figure}[t]
	\includegraphics[width=3.4in]{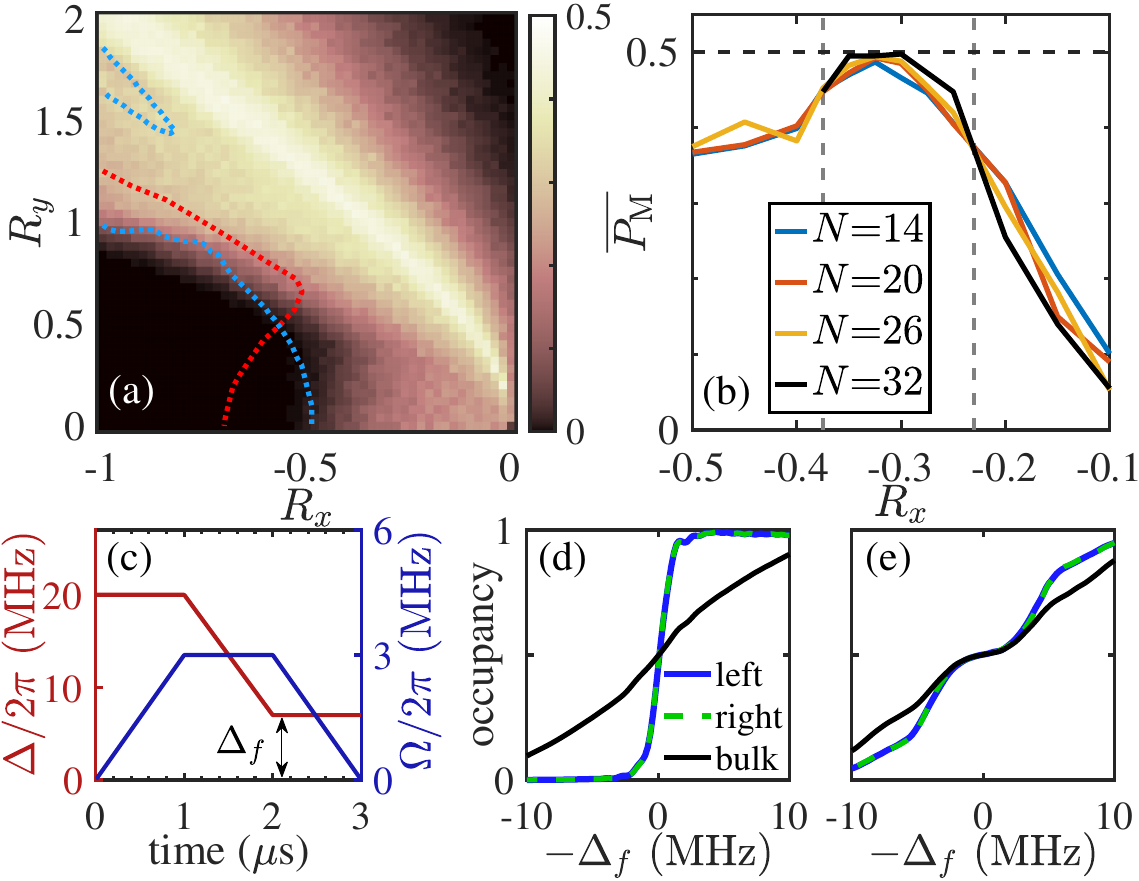}
	\caption{(Color online) 
		(a) $\mathbb{Z}_2$ invariant $\overline{P_\mathrm{M}}$ versus $R_x$ and $R_y$ for an amorphous lattice with $N=8$.
		For comparison, we also plot the phase boundaries for hard-core bosons at half-filling in a regular lattice as cyan dotted lines and the phase boundary in
		the single-particle level for an amorphous lattice with size $N=8$ as a red dotted line. 
		(b) $\overline{P_\mathrm{M}}$ versus $R_x$ for various system sizes when $R_y=1$. 
			(c) Microwave sweep with the Rabi frequency $\Omega$ and detuning $\Delta$ varying with time; $\Delta$ ends at $\Delta_f$.
			(d) and (e) Bulk (black curves) and edge (blue and dashed green curves) sites occupancy after a microwave sweep with respect to the final detuning $\Delta_f$ 
			in a topological phase with $R_x=-0.33$ and $R_y=1$ and in a trivial phase with $R_x=0.45$ and $R_y=1$, respectively. In (c-e), $N=7$.
		Here, $R_z=0.7$. 
	}
	\label{fig3}
\end{figure}

If we consider Hamiltonian (\ref{Hmodel}) with only NN hopping, we can show that the Hamiltonian can support a topological amorphous phase using a similar argument as that in the single-particle case~\cite{Suppl}.
Such a Hamiltonian can also be mapped to a free fermionic model~\cite{Suppl} by the inverse Jordan-Wigner transformation, $\hat{b}_i^\dagger = [\prod_{j=1}^{i-1} (1-2\hat{c}_j^\dagger\hat{c}_j)] \hat{c}_i^\dagger$, where $\hat{c}_i^\dagger$ is a fermion creation operator at site $i$.
However, with long-range hopping, we can no longer map $\hat{H}$ to a free fermionic Hamiltonian~\cite{Browaeys2019Science,Suppl}.
In this case, the system becomes a true many-body system with interactions. We will use the ED for $N\leq 10$ and MPS for $N>10$ to calculate the ground state of the Rydberg Hamiltonian (\ref{Hmodel}).

To characterize the topological property of the many-body system, we define a $\mathbb{Z}_2$ invariant as~\cite{VBS2002,Tasaki2018PRL}
\begin{equation}
	P_\mathrm{M} = \frac{1}{2\pi} \mathrm{Im} \ln 
	\bra{\Psi_0} \hat{\mathcal{P}}_\mathrm{M} \ket{\Psi_0},
\end{equation}
where $\ket{\Psi_0}$ is the many-body ground state of the hard-core bosonic Hamiltonian for periodic boundaries, and $\hat{\mathcal{P}}_\mathrm{M} = \prod_{j=1}^{2N} e^{- \frac{\pi i}{N} x_j \sigma_{j}^z }$ is the twist operator~\cite{footnote1}. 
Hamiltonian~(\ref{Hmodel}) respects the particle-hole symmetry, i.e., $\hat{\Xi} \hat{H} \hat{\Xi}^{-1}=\hat{H}$ with $\hat{\Xi}=\prod_{j=1}^{2N} \sigma_j^x$, the time-reversal symmetry, i.e., $\hat{T}\hat{H}\hat{T}^{-1}=\hat{H}$ with $\hat{T}=\prod_{j=1}^{2N}\sigma_j^y \kappa$, and an anti-unitary symmetry, i.e., $\hat{\mathcal{S}} \hat{H} \hat{\mathcal{S}}^{-1}=\hat{H}$ with $\hat{\mathcal{S}}=\prod_{j=1}^{2N} \sigma_j^x \kappa$ and $\kappa$ being the complex conjugate operator.
All these symmetries can protect the quantization of the $\mathbb{Z}_2$ invariant for a many-body eigenstate $\ket{\Phi}$ that is not degenerate for periodic boundaries~\cite{footnote2}.
Indeed, our numerical results show that the ground states obtained by the ED have quantized values for $P_\mathrm{M}$ for each sample. 
For those calculated by the MPS, their $P_\mathrm{M}$ are very close to be quantized, which is reasonable given that the MPS can only find approximate ground states.

From the phase diagram in Fig.~\ref{fig3}(a), we see a long narrow region with $\overline{P_\mathrm{M}}$ close to $0.5$, signaling the existence of a topological amorphous phase for hard-core bosons at half-filling. 
The figure also illustrates the existence of a topologically trivial region with zero $\overline{P_\mathrm{M}}$ and a large intermediate region with $0<\overline{P_\mathrm{M}}<0.5$ (due to the coexistence of trivial and nontrivial samples) between these two phases.
Compared with the single-particle case in amorphous lattices and the many-body case in regular lattices, whose phase boundaries are shown by dotted lines, the parameter region with nonzero $\overline{P_\mathrm{M}}$ shrinks in large parts.
Interestingly, there exists a region around $R_y=1.75$ where the phase is trivial for a regular lattice while the phase is in an intermediate region for an amorphous lattice. 
We note that whether this indicates that structural disorder can induce a topological phase transition in the many-body case is still unclear due to the system size limitation. 

To further identify the existence of a topological phase in the many-body case, we show $\overline{P_\mathrm{M}}$ versus $R_x$ for distinct system sizes in Fig.~\ref{fig3}(b). 
We see that there exists a parameter region for $-0.375 \lesssim R_x \lesssim -0.23$ where $\overline{P_\mathrm{M}}$ is approaching $0.5$ as the system size $N$ increases. 
One can also find the finite-size analysis in the Supplementary Material, which further illustrates that the regime is topologically nontrivial.
In other regions such as $R_x > -0.23$, $\overline{P_\mathrm{M}}$ declines as $N$ increases, suggesting a trivial phase. 
In the region for $-0.5<R_x <-0.375$, current numerical results suggest that it is an intermediate region. 
But the conclusion may change for larger system sizes. 
In addition, our numerical results suggest that all these phases are gapless~\cite{Suppl}.

To experimentally observe the topological phase in the many-body case, one can shine a global microwave radiation 
with time-varying Rabi frequency and detuning ending at $\Delta_f$ [see Fig.~\ref{fig3}(c)], similar to the experiment in Ref.~\cite{Browaeys2019Science}.
Now, $-\Delta_f$ plays the role of the chemical potential for the Hamiltonian $-\hat{H}$, which is topologically equivalent to $\hat{H}$~\cite{Suppl}. 
We numerically simulate the full time evolution, and the results show that in the topological amorphous phase, as $-\Delta_f$ changes from negative to positive values across zero, the bulk sites occupancy increases continuously without experiencing a plateau, revealing the gapless property of the system. However, the edge sites occupancy exhibits a sharp rise across zero detuning, indicating the emergence of particles localized at the edges. For comparison, we also present the results in the trivial phase where the sharp rise is not observed.

In addition, in the topological phase, we find that the invariant $\overline{P_\mathrm{M}}=0.37$ ($0.35$), density-density correlations $\overline{C^z}_{\textrm{inter}}=\sum_{i=1}^{N-1} \overline{C_{2i,2i+1}^z} / (N-1)=-0.44$ ($-0.48$) 
and $\overline{C^z}_{\textrm{intra}}=\sum_{i=1}^N \overline{C_{2i-1,2i}^z} / N=-0.08$ ($-0.10$), 
and the string order parameter $\overline{C^z}_{\textrm{string}}=(-1)^{N-1} \overline{ \langle \prod_{i=2}^{2N-1} \sigma_i^z \rangle }=0.25$ ($0.65$) for the prepared state after the sweep for $\Delta_f=0$;
the results are consistent with those of the topological ground state
displayed in the bracket.
Here, $C_{i,j}^z=\langle \sigma_i^z \sigma_j^z \rangle - \langle \sigma_i^z \rangle \langle \sigma_j^z \rangle$. In the trivial phase, we obtain $\overline{P_\mathrm{M}}=0.005$ ($0.000$), 
$\overline{C^z}_{\textrm{inter}}=-0.21$ ($-0.22$) 
and $\overline{C^z}_{\textrm{intra}}=-0.47$ ($-0.49$),
and $\overline{C^z}_{\textrm{string}}=0.0015$ ($0.0017$). All these quantities are experimentally accessible~\cite{Browaeys2019Science}.

In summary, we have predicted the existence of topological amorphous phases in an experimentally accessible 
Rydberg chain in both single-particle and many-body levels. In the single-particle level, 
we also find a structural disorder induced topological phase transition. 
Our numerical simulations of the time evolution further provide strong evidence that these interesting phenomena can be experimentally 
observed using state-of-the-art technologies.

\begin{acknowledgments}
We thank D.-L. Deng and Y.-L. Tao for helpful discussion. The work is
supported by the National Natural Science Foundation
of China (11974201), the start-up fund from Tsinghua University and
the National Thousand-Young-Talents Program.
\end{acknowledgments}

\begin{widetext}
	\setcounter{equation}{0} \setcounter{figure}{0} \setcounter{table}{0} %
	\renewcommand{\theequation}{S\arabic{equation}} \renewcommand{\thefigure}{S%
		\arabic{figure}}
	\renewcommand{\bibnumfmt}[1]{[S#1]} \renewcommand{%
	\citenumfont}[1]{S#1}

	In the supplementary material, we will 
	present the energy gap and the level-spacing ratio (LSR) for the single-particle Hamiltonian in Section S-1, 
	prove the quantization of the polarization protected by chiral symmetry in Section S-2, 
	show the results using other types of distributions for structural disorder in Section S-3,
	discuss the existence of topological amorphous phases in the Hamiltonian with only nearest-neighbor (NN) hopping in the many-body level in Section S-4,
	explain the many-body effects in the Hamiltonian with long-range hopping in Section S-5, 
	prove the quantization of a $\mathbb{Z}_2$ invariant for an arbitrary spin system in Section S-6, 
	discuss the property of the ground states in the many-body case at half-filling in Section S-7,
	present the finite size analysis, the energy gap, and the structural disorder induced topological phase transition for the many-body Hamiltonian in Section S-8, 
	and finally present the details on numerical simulations of experimental observations of topological phases in Section S-9.
	
	\section{S-1. Energy gap and LSR in the single-particle case}
	In this section, we will show that the topological amorphous phases in the single-particle case are gapless and localized.
	Specifically, we plot the energy gap around zero energy in Fig.~\ref{figS1}(a), showing the gapless feature.
	
	To characterize the localization property, we define the LSR for all states with negative energies containing $N_E$ energy levels as
	\begin{equation}
		\overline{r}=\left[\frac{1}{N_E-2}\sum_{n=1}^{N_E-2} \frac{\min \{\delta_n , \delta_{n+1}\}}{\max \{\delta_n , \delta_{n+1}\}} \right],
	\end{equation}
	where $\delta_n = E_{n+1} - E_n$ is the energy difference between two consecutive energy levels $E_n$ and $E_{n+1}$
	(here we assort the energy levels in ascending order), and $[\cdots]$ denotes the average over different disorder realizations.
	For our system, when the states are localized, $\overline{r} \approx 0.386$ corresponding to the Poisson statistics, 
	and when the states are extended, $\overline{r} \approx 0.53$ corresponding to the Gaussian orthogonal ensemble (GOE)~\cite{Huse2007PRBS}.
	Our calculation shows that the LSR is close to $0.386$, indicating that all 
	states are localized in the parameter region as shown in Fig.~\ref{figS1}(b).
	
	\begin{figure*}[t]
		\includegraphics[width=4.7in]{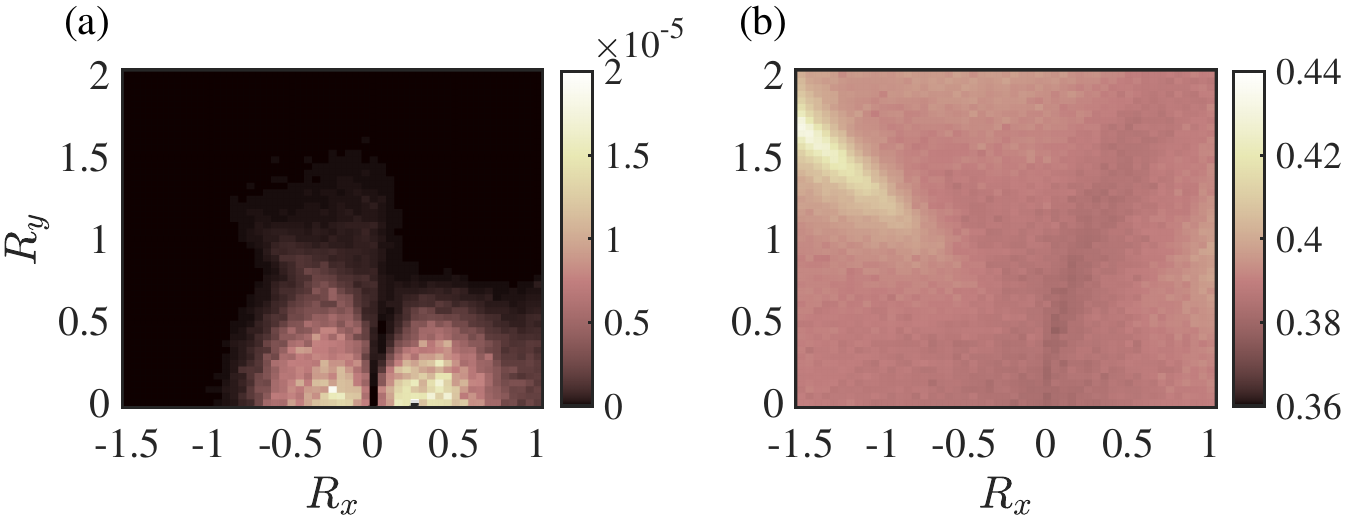}
		\caption{(Color online) 
			(a) The sample averaged energy gap around zero energy and (b) the sample averaged LSR with respect to $R_x$ and $R_y$ for the single-particle Hamiltonian $H^{\mathrm{S}}$. 
			The LSR is calculated based on all negative energies.
			Here, $R_z=0.8$, the system size $N=200$, and all quantities are averaged over $500$
			random configurations.
		}
		\label{figS1}
	\end{figure*}

	\section{S-2. Proof of the quantization of the polarization protected by chiral symmetry}
	We now prove that the quantization of the polarization $P_\mathrm{S}$ can be protected by chiral symmetry.
	We begin by defining the polarization $P_\mathrm{S}^\mathrm{u}$ as 
	\begin{equation}
		P_\mathrm{S}^\mathrm{u} = [\frac{1}{2\pi} \mathrm{Im} \ln \det (U_\mathrm{u}^\dagger D U_\mathrm{u})
		-  \frac{1}{2N} \sum_{i =1}^{2N} x_i ]\mod 1,
		\label{polar_u}
	\end{equation}
	where $U_\mathrm{u}=\Pi U$ is a $2N\times N$ matrix composed of all eigenstates of $H^\mathrm{S}$ with positive energies.
	From
	\begin{equation}
		\det (U_\mathrm{u} ^\dagger D U_\mathrm{u}) = 
		\det (U^\dagger \Pi D \Pi U) = \det (U^\dagger D \Pi \Pi U) = \det (U^\dagger D  U),
	\end{equation}
	where we have used $\Pi D=D\Pi$ and $\Pi^2=\mathbf{1}_{2N}$, it can be easily confirmed that 
	\begin{equation}
		P_\mathrm{S}=P_\mathrm{S}^\mathrm{u}.
		\label{E1}
	\end{equation}
	
	When all states are occupied, we can define the polarization as
	\begin{equation}
		P_\mathrm{S}^\mathrm{f} = \left[\frac{1}{2\pi} \mathrm{Im} \ln \det (U_\mathrm{f}^\dagger D U_\mathrm{f})
		-  \frac{1}{N} \sum_{i =1}^{2N} X_i\right] \mod 1,
		\label{Spolar_full_SM}
	\end{equation}
	where $U_\mathrm{f}= (U ~ U_\mathrm{u})$ is a $2N\times 2N$ unitary matrix composed of all the eigenstates of $H^\mathrm{S}$. 
	Since
	\begin{equation}
		\det (U_\mathrm{f}^\dagger D U_\mathrm{f}) = \det (D U_\mathrm{f} U_\mathrm{f}^\dagger)
		= \det(D)=\prod_{\alpha=1}^{2N} e^{2 \pi i X_\alpha /N},
	\end{equation}
	we have
	\begin{equation}
		P_\mathrm{S}^\mathrm{f}=\left[\frac{1}{2\pi} \mathrm{Im} 
		\ln  (\prod_{\alpha=1}^{2N} e^{2 \pi i X_\alpha /N})
		-  \frac{1}{N} \sum_{i =1}^{2N} X_i\right] \mod 1 ~ = 0.
		\label{E2}
	\end{equation}
	
	We can also prove that
	\begin{equation}
		P_\mathrm{S}^\mathrm{f} = (P_\mathrm{S} + P_\mathrm{S}^\mathrm{u}) \mod 1,
		\label{E3}
	\end{equation}
	because
	\begin{equation}
		\begin{aligned}
			&(P_\mathrm{S}^\mathrm{f} - P_\mathrm{S} - P_\mathrm{S}^\mathrm{u}) \mod 1 \\
			=& \left[
			\frac{1}{2\pi} \mathrm{Im} \ln \det (U_\mathrm{f}^\dagger D U_\mathrm{f}) 
			- \frac{1}{2\pi} \mathrm{Im} \ln \det (U^\dagger D U)
			- \frac{1}{2\pi} \mathrm{Im} \ln \det (U_\mathrm{u}^\dagger D U_\mathrm{u})\right] \mod 1\\
			=& \left[\frac{1}{2\pi} \mathrm{Im} \ln \det 
			\begin{pmatrix}
				U^\dagger D U            & U^\dagger D U_\mathrm{u}            \\
				U_\mathrm{u}^\dagger D U & U_\mathrm{u}^\dagger D U_\mathrm{u}
			\end{pmatrix}
			+ \frac{1}{2\pi} \mathrm{Im} \ln \det (U^\dagger D^\dagger U)
			+ \frac{1}{2\pi} \mathrm{Im} \ln \det (U_\mathrm{u}^\dagger D^\dagger U_\mathrm{u}) \right]\mod 1\\
			=& \left[\frac{1}{2\pi} \mathrm{Im} \ln \det 
			\begin{pmatrix}
				U^\dagger D U            & U^\dagger D U_\mathrm{u}            \\
				U_\mathrm{u}^\dagger D U & U_\mathrm{u}^\dagger D U_\mathrm{u}
			\end{pmatrix}
			+ \frac{1}{2\pi} \mathrm{Im} \ln \det 
			\begin{pmatrix}
				U^\dagger D^\dagger U & U^\dagger D^\dagger U_\mathrm{u}            \\
				0                     & U_\mathrm{u}^\dagger D^\dagger U_\mathrm{u}
			\end{pmatrix} \right]\mod 1 \\
			=& \left[\frac{1}{2\pi} \mathrm{Im} \ln \det 
			\begin{bmatrix}
				\begin{pmatrix}
					U^\dagger D U            & U^\dagger D U_\mathrm{u}            \\
					U_\mathrm{u}^\dagger D U & U_\mathrm{u}^\dagger D U_\mathrm{u}
				\end{pmatrix}
				\begin{pmatrix}
					U^\dagger D^\dagger U & U^\dagger D^\dagger U_\mathrm{u}            \\
					0                     & U_\mathrm{u}^\dagger D^\dagger U_\mathrm{u}
				\end{pmatrix}
			\end{bmatrix}\right] \mod 1\\
			=& \left[\frac{1}{2\pi} \mathrm{Im} \ln \det 
			\begin{pmatrix}
				(U^\dagger D U)(U^\dagger D U)^\dagger         & 0                                 \\
				U_\mathrm{u}^\dagger D U U^\dagger D^\dagger U & U_\mathrm{u}^\dagger U_\mathrm{u}
			\end{pmatrix} \right]\mod 1\\
			=& \left[\frac{1}{2\pi} \mathrm{Im} \ln \det ((U^\dagger D U)(U^\dagger D U)^\dagger)\right] \mod 1\\
			=& 0,
		\end{aligned}
	\end{equation}
	where we have used $U^\dagger U = U_\mathrm{u}^\dagger U_\mathrm{u} = \mathbf{1}_{N}$, 
	$U^\dagger U_\mathrm{u}=0$ and 
	$U U^\dagger + U_\mathrm{u} U_\mathrm{u}^\dagger = \mathbf{1}_{2N}$. Based on Eqs.~(\ref{E1},\ref{E2},\ref{E3}),
	we conclude that $P_\mathrm{S}$ can only take discrete values of $0$ or $0.5$ due to the presence of chiral symmetry.
	
	\section{S-3. other types of structural disorder}
	
	\begin{figure*}
		\includegraphics[width=5in]{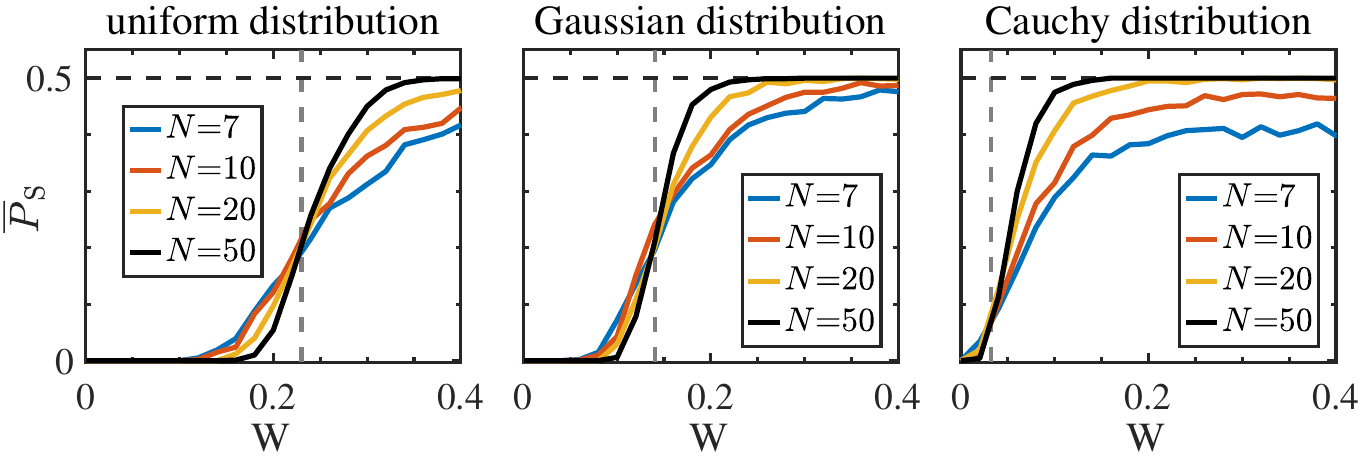}
		\caption{(Color online) 
			Polarization $\overline{P_\mathrm{S}}$ as a function of the disorder strength $W$ for three different types of structural disorder. 
			The disorder strength $W$ is twice the standard deviation for the Gaussian distribution and is the full width at half maximum for the Cauchy distribution. 
			The system parameters are the same as those in Fig.~2(d) in the main text, namely, $R_x=-0.5$, $R_y=0$ and $R_z=0.8$.    
		}
		\label{figS5}
	\end{figure*}

	In the main text, we consider structural disorder from a uniform distribution. In this section, 
	we consider other types of distributions including the Gaussian distribution $f_G(\delta z_i)=\frac{2}{W\sqrt{2\pi}} e^{-2(\delta z_i)^2/W^2}$ with 
	$W/2$ being the standard deviation
	and the Cauchy distribution $f_C(\delta z_i)=\frac{2}{\pi W} \cdot \frac{(W/2)^2}{(\delta z_i)^2 + (W/2)^2}$ with $W$ being the full width at half maximum.
	We find that the structural disorder induced topological phase transition occurs for all the three types of distributions but at different disorder strength $W$, 
	as shown in Fig.~\ref{figS5}. The results imply that the phase transition is not restricted to specific structural disorder distributions.
	
	\section{S-4. Topological amorphous phases in the Hamiltonian with the nearest-neighbor (NN) hopping in the many-body level}
	\subsection{A. The existence of many-body topological amorphous phases}
	In the main text, we have argued that in the single-particle case, the NN Hamiltonian can host topological phases in a random geometry, given
	the fact that there exists a parameter region where the hopping within a unit cell vanishes so that two zero-energy edge states appear under open boundary conditions.
	In the many-body case, we can still consider the scenario with vanishing hopping within a unit cell.
	In this case, even though the unit cells are randomly distributed, the first and final sites do not couple to the other sites
	under open boundary conditions. As a consequence, $\hat{b}_1$, $\hat{b}_1^\dagger$, $\hat{b}_{2N}$ and $\hat{b}_{2N}^\dagger$ do not exist in the NN Hamiltonian and thus they commute with the Hamiltonian. This leads to four degenerate many-body ground states under open boundary conditions: $|\phi_0\rangle$, $\hat{b}_1^\dagger |\phi_0\rangle$, $\hat{b}_{2N}^\dagger|\phi_0\rangle$ 
	and $\hat{b}_1^\dagger \hat{b}_{2N}^\dagger |\phi_0\rangle$, revealing the topological property of the many-body system.
	
	\subsection{B. The relation between the free fermionic Hamiltonian and the hard-core bosonic Hamiltonian by the Jordan-Wigner transformation}
	In this subsection, we will show that for a system with odd number of unit cells, the free fermionic Hamiltonian 
	\begin{equation}
		\hat{H}_F=\sum_{i=1}^{2N-1} V_{i,i+1}(\hat{c}_i^\dagger \hat{c}_{i+1}+H.c.)+
		V_{1,2N}\hat{c}_1^\dagger \hat{c}_{2N}+V_{1,2N}\hat{c}_{2N}^\dagger \hat{c}_1
	\end{equation}
	with the NN hopping under periodic boundary conditions at half-filling 
	can be transformed to 
	the hard-core bosonic model with the NN hopping by the Jordan-Wigner transformation,
	\begin{equation}
		\begin{aligned}
			\hat{c}_i^\dagger &= (\prod_{\alpha=1}^{i-1} 
			\sigma^z_{\alpha}) \cdot b_i^\dagger,\\
			\hat{c}_i  &= (\prod_{\alpha=1}^{i-1} \sigma^z_{\alpha}) \cdot \hat{b}_i.
		\end{aligned}
	\end{equation}
	From the transformation, one can easily find that $\hat{c}_i^\dagger \hat{c}_{i+1}=\hat{b}_i^\dagger \hat{b}_{i+1}$ when $i<2N$. For the hopping term from the initial site to the end one or vice versa, 
	we have
	\begin{equation}
		\hat{c}_1^\dagger \hat{c}_{2N}=\hat{b}_1^\dagger \cdot (\prod_{\alpha=1}^{2N-1} \sigma_\alpha^z) \cdot \hat{b}_{2N}
		= - \hat{b}_1^\dagger \hat{b}_{2N} \cdot (\prod_{\alpha=1}^{2N} \sigma_\alpha^z),
	\end{equation}
	and
	\begin{equation}
		\hat{c}_{2N}^\dagger \hat{c}_1=(\prod_{\alpha=1}^{2N-1} \sigma_\alpha^z) \cdot \hat{b}_{2N}^\dagger \hat{b}_1
		= - \hat{b}_{2N}^\dagger \hat{b}_1 \cdot (\prod_{\alpha=1}^{2N} \sigma_\alpha^z).
	\end{equation}
	We can always consider a basis consisting of Fock states with fixed total particle numbers since the total particle number is conserved, 
	that is, $\beta=\{|n_1 n_2 \cdots n_{2N}\rangle\}$ where
	$|n_1 n_2 \cdots n_{2N}\rangle$ is a Fock state with $n_j$ hard-core bosons at site $j$ with $j=1,2,\cdots, 2N$ ($n_j$ can take the value of either $0$ or $1$).
	At half-filling, $\sum_{j=1}^{2N}n_j=N$. We therefore have
	$(\prod_{\alpha=1}^{2N} \sigma_\alpha^z)|n_1 n_2 \cdots n_{2N}\rangle=\prod_{\alpha=1}^{2N}(1-2\hat{b}_\alpha^\dagger\hat{b}_\alpha)|n_1 n_2 \cdots n_{2N}\rangle=
	(-1)^{N} |n_1 n_2 \cdots n_{2N}\rangle$,
	indicating that when $N$ is odd (i.e., there are odd number of unit cells), $\hat{c}_1^\dagger \hat{c}_{2N}=\hat{b}_1^\dagger \hat{b}_{2N}$ and 
	$\hat{c}_{2N}^\dagger \hat{c}_1=\hat{b}_{2N}^\dagger \hat{b}_1 $. We thus conclude that for a system with odd number of unit cells, 
	the free fermionic Hamiltonian with the NN hopping under periodic boundary conditions
	at half-filling
	can be transformed to the hard-core bosonic model with the NN hopping, i.e., 
	\begin{equation}
		\hat{H}_F[\hat{c}_i^\dagger,\hat{c}_i]=\hat{H}_B[\hat{b}_i^\dagger,\hat{b}_i],
	\end{equation}
	where 
	\begin{equation}
		\hat{H}_B=\sum_{i=1}^{2N-1} V_{i,i+1}(\hat{b}_i^\dagger \hat{b}_{i+1}+H.c.)+
		V_{1,2N}\hat{b}_1^\dagger \hat{b}_{2N}+V_{1,2N}\hat{b}_{2N}^\dagger \hat{b}_1.
	\end{equation}
	
	\subsection{C. The topological equivalence between the free fermionic Hamiltonian and the hard-core bosonic Hamiltonian}
	We now show that for a system with odd number of unit cells, the $\mathbb{Z}_2$ index of the NN hard-core bosonic model is equal to the polarization of a NN free fermionic model.
	Let $\ket{\Phi_0}$ be the many-body ground state of 
	$\hat{H}_F$, which is a Slater determinant of all the single-particle eigenstates of $\hat{H}_F$ with negative energies. 
	Since $\hat{H}_B [\hat{b}_i^\dagger,\hat{b}_i]=\hat{H}_F [\hat{c}_i^\dagger,\hat{c}_i]$,
	$\ket{\Phi_0}$ is in fact also the many-body ground state of $\hat{H}_B$. 
	We also note that the polarization $P_S$ defined in Eq. (2) in the main text can also be written as  $P_\mathrm{S}=\frac{1}{2\pi} \mathrm{Im} \ln \bra{\Phi_0} \hat{\mathcal{P}}_\mathrm{S} \ket{\Phi_0}$ for a free fermionic model
	where $\hat{\mathcal{P}}_\mathrm{S} =  e^{\frac{2 \pi i}{N}  \sum_{j=1}^{2N} x_{j} (\hat{c}_{j}^\dagger \hat{c}_{j} -\frac{1}{2}) }$.
	By the Jordan-Wigner transformation, 
	$\hat{\mathcal{P}}_\mathrm{S}=\hat{\mathcal{P}}_\mathrm{M}$, indicating that the $\mathbb{Z}_2$ index 
	\begin{equation}
		P_\mathrm{M} = \frac{1}{2\pi} \mathrm{Im} \ln \bra{\Phi_0} \hat{\mathcal{P}}_\mathrm{M} \ket{\Phi_0} = \frac{1}{2\pi} \mathrm{Im} \ln \bra{\Phi_0} \hat{\mathcal{P}}_\mathrm{S} \ket{\Phi_0} = P_\mathrm{S}.
	\end{equation}
	Therefore, 
	$\hat{H}_B$ and $\hat{H}_F$ share the same topology. We have also numerically confirmed that the topological invariants for $\hat{H}_B$ and $\hat{H}_F$ are indeed equal
	for all samples when there are odd number of unit cells. While the proof based on the Jordan-Wigner transformation is restricted to a system with odd number of unit cells,
	the argument in subsection A can be applied to both even and odd cases. All these results indicate the existence of many-body topological amorphous phases in the NN hard-core bosonic Hamiltonian. 
	
	\section{S-5. The many-body Hamiltonian with long-range interactions}
	In this section, we show that with long-range hoppings, the hard-core bosonic model can be mapped into a fermionic model with
	interactions (see also Ref.~\cite{Browaeys2019ScienceS}). For clarity, we consider a simpler Hamiltonian
	with up to next-next-nearest-neighbor hopping,
	\begin{equation}
		\hat{H} = \sum_i ( V_{i,i+1} \hat{b}_i^\dagger \hat{b}_{i+1} + V_{i,i+3} \hat{b}_i^\dagger b_{i+3} + H.c.),
	\end{equation}
	which can be mapped to a fermionic model via Jordan-Wigner transformation,
	\begin{equation}
		\hat{H} = 
		\sum_i \left[ V_{i,i+1} \hat{c}_i^\dagger \hat{c}_{i+1} + V_{i,i+3} \hat{c}_i^\dagger \hat{c}_{i+3}+
		2V_{i,i+3} 
		\hat{c}_i^\dagger (2\hat{n}_{i+1}\hat{n}_{i+2}-\hat{n}_{i+1}-\hat{n}_{i+2} ) \hat{c}_{i+3} + H.c.\right], 
	\end{equation}
	with $\hat{c}_i^\dagger$ ($\hat{c}_i$) being a fermionic creation (annihilation) operator at site $i$ and $\hat{n}_i=\hat{c}_{i}^\dagger \hat{c}_i$ being a particle number operator.
	Without $V_{i,i+3}$, the Hamiltonian corresponds to a non-interacting fermionic model. However, in the presence of these terms, 
	the corresponding fermionic model contains interactions besides the non-interacting hopping terms, implying that 
	the system is a genuine interacting system.
	
	\section{S-6. Quantization of the $\mathbb{Z}_2$ invariant for an arbitrary spin system}
	Ref.~\cite{VBS2002S} proves that the $\mathbb{Z}_2$ invariant for a spin-1 system is quantized due to 
	time-reversal and spin-rotational symmetries. We now generalize the results to an arbitrary spin system,
	showing that the time-reversal symmetry, the spin-rotational symmetries and other two anti-unitary symmetries can protect
	the quantization of the $\mathbb{Z}_2$ invariant defined as
	\begin{equation}
		\mathcal{P}_A = \frac{1}{2\pi} \mathrm{Im} \ln 
		\bra{\Psi_0} \hat{\mathcal{P}}_A \ket{\Psi_0},
	\end{equation}
	where $\ket{\Psi_0}$ is the ground state and $\hat{\mathcal{P}}_A=\prod_{j=1}^{M} e^{- 2\pi i 
		x_j S_{j}^z/L }$ with $S_j^{z}$ being a spin operator along $z$ at site $j$, $M$ being the total number
	of spins and $L$ being the length of the system.
	
	Suppose that the many-body ground state $\ket{\Psi_0}$ of a Hamiltonian $\hat{H}$ under periodic boundary condition is not degenerate.
	When a system respects the time-reversal symmetry $\hat{T}_A=\prod_{j=1}^{M}e^{-i\pi S_j^y}\kappa$ with $\hat{T}_A^2=1$ for even $M$, we have $\hat{T}_A\ket{\Psi_0}=e^{i\theta} \ket{\Psi_0}$.
	We can thus derive
	\begin{equation}
		\bra{\Psi_0} \hat{\mathcal{P}}_A \ket{\Psi_0}
		= \bra{\hat{T}_A \Psi_0} \hat{\mathcal{P}}_A \ket{\hat{T}_A \Psi_0}
		= \langle {\hat{T}_A \Psi_0} | \hat{T}_A \hat{\mathcal{P}}_A { \Psi_0} \rangle 
		= \langle \Psi_0 | \hat{\mathcal{P}}_A | \Psi_0 \rangle ^ *,
	\end{equation}
	where we have used the result that
	$\hat{T}_A \hat{\mathcal{P}}_A \hat{T}_A^{-1} = \hat{\mathcal{P}}_A $. 
	Since $\bra{\Psi_0} \hat{\mathcal{P}}_\mathrm{M} \ket{\Psi_0}$ is real, $\mathcal{P}_A$ for the state $\ket{\Psi_0}$ can only take discrete values of
	zero or $0.5$ up to an integer. 
	
	We now consider two spin-rotational symmetries: $\hat{R}_x=\prod_{j=1}^{M}e^{-i\pi S_j^x}$ and $\hat{R}_y=\prod_{j=1}^{M}e^{-i\pi S_j^y}$.
	Similarly, we have 
	$\hat{R}_\nu \ket{\Psi_0} = \pm \ket{\Psi_0}$ ($\nu=x,y$) since $\hat{R}_\nu^2=1$ for an even $M$.
	Note that for a half spin system, $\hat{R}_x$ is also known as the particle-hole symmetry. One can easily derive that 
	$\hat{R}_\nu^{\dagger} \hat{\mathcal{P}}_\mathrm{M} \hat{R}_\nu = \hat{\mathcal{P}}_\mathrm{M}^\dagger $ ($\nu=x,y$),
	which leads to
	\begin{equation}
		\bra{\Psi_0} \hat{\mathcal{P}}_A \ket{\Psi_0}
		= \bra{ \hat{R}_\nu \Psi_0} \hat{\mathcal{P}}_A  \ket{ \hat{R}_\nu \Psi_0}
		= \langle {\Psi_0} | \hat{R}_\nu^{\dagger} \hat{\mathcal{P}}_A  \hat{R}_\nu | {\Psi_0} \rangle 
		= \langle \Psi_0 | \hat{\mathcal{P}}_A ^\dagger | \Psi_0 \rangle 
		= \langle \Psi_0 | \hat{\mathcal{P}}_A  | \Psi_0 \rangle ^*.
	\end{equation}
	Therefore, $\mathcal{P}_A$ can only take discrete values of $0$ or $0.5$ up to an integer.
	
	We can also consider other anti-unitary symmetries such as 
	$\hat{S}_A= \hat{R}_\nu \kappa$ with $\nu=x,y$. We also can derive that $\hat{S}_A \hat{\mathcal{P}}_A \hat{S}_A^{-1}=\hat{\mathcal{P}}_A$, leading to 
	\begin{equation}
		\bra{\Psi_0} \hat{\mathcal{P}}_A \ket{\Psi_0}
		= \bra{\hat{S}_A \Psi_0} \hat{\mathcal{P}}_A \ket{\hat{S}_A \Psi_0}
		= \langle {\hat{S}_A \Psi_0} | \hat{S}_A \hat{\mathcal{P}}_A { \Psi_0} \rangle 
		= \langle \Psi_0 | \hat{\mathcal{P}}_A | \Psi_0 \rangle ^ *,
	\end{equation}
	so that $\mathcal{P}_A$ has to take discrete values of $0$ or $0.5$ up to an integer.
	
	To sum up, for a system consisting of even number of arbitrary spins, the $\mathbb{Z}_2$ invariant is 
	enforced to be quantized by the time-reversal, spin-rotational and other anti-unitary symmetries.
	
	\section{S-7. Discussion on the property of the ground state in the many-body case at half-filling}
	Our spin model also respects a $U(1)$ symmetry, i.e., $[\hat{H},s^z]$ with $s^z=\sum_{j=1}^{2N}\sigma_j^z/2$ being the total spin operator along $z$ so that $s^z$ is a conserved quantity. 
	If a many-body eigenstate $\ket{\phi}$ of $\hat{H}$ has nonzero eigenvalues $m_z$ ($m_z \neq 0$) of $s^z$, then $\hat{S}\ket{\phi}$ must be another eigenstate with the same energy as $\ket{\phi}$ but opposite $s^z$ eigenvalue since $\{\hat{S},s^z\}=0$, implying that $\ket{\phi}$ is degenerate. 
	Thus, a nondegenerate state, if it exists, must lie in the subspace with $m_z=0$. 
	In fact, such a condition corresponds to the constraint of half-filling for the hard-core bosonic model. 
	It is a well-known fact that, without magnetic fields, the ground state for the XY spin chain with only short-range coupling in regular lattices has a zero total spin, and the state is not degenerate for periodic boundaries. 
	These properties should remain for the model including long-range couplings.
	As a result, $P_\mathrm{M}$ can only take quantized values for the ground state and thus can be used as a topological invariant. 
	In the amorphous case, we expect that these properties remain unchanged.
	Indeed, our numerical results show that the ground states obtained by the ED have $m_z=0$ and quantized values for $P_\mathrm{M}$ for each sample. 
	For those calculated by the MPS, their $P_\mathrm{M}$ are very close to be quantized, which is reasonable given that the MPS can only find approximate ground states.
	
	\section{S-8. Finite size analysis, energy gap, and structural disorder induced topological phase transition for the many-body Hamiltonian}
	
	\begin{figure*}[t]
		\includegraphics[width=4in]{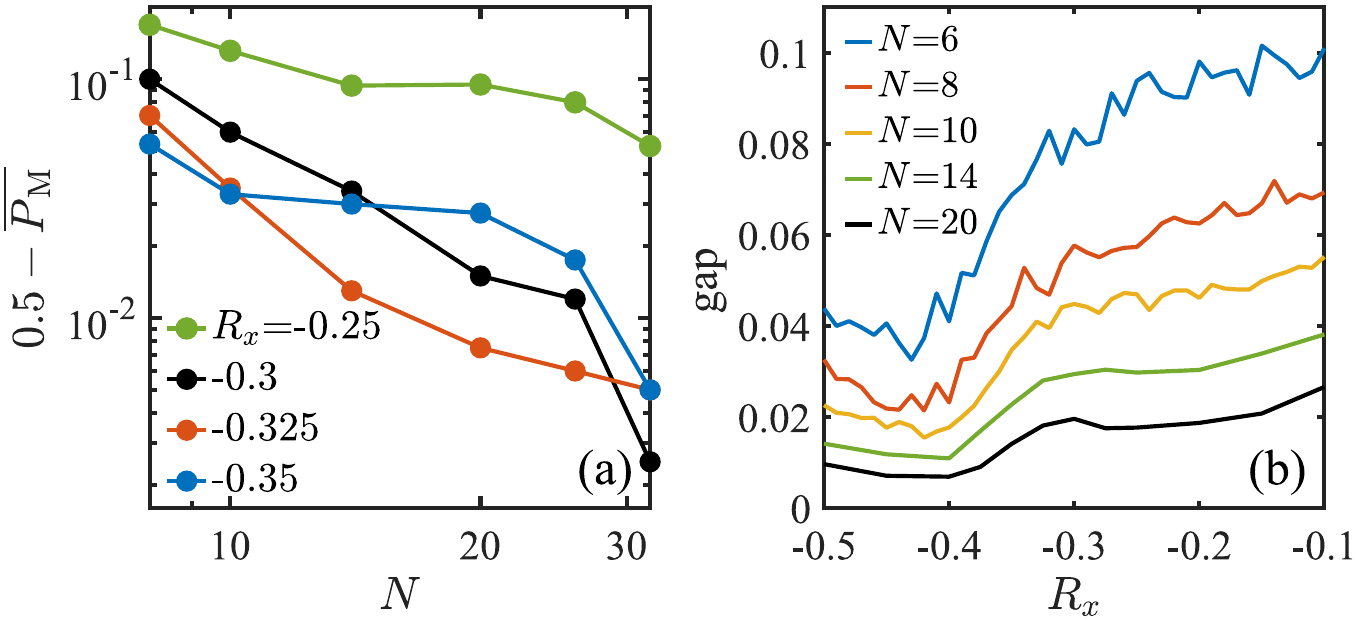}
		\caption{(Color online) 
			(a) $0.5-\overline{P_\mathrm{M}}$ with respect to the system size $N$ for different $R_x$ in the logarithmic scale.
			(b) The sample averaged energy gap of the many-body Hamiltonian with respect to $R_x$ for different system sizes.		
			The energy of the ground state and the first excited state is calculated via exact 
			diagonalization when $N \le 10$ and matrix product states when $N > 10$. 
			The average is performed over more than $200$ samples.
			Here $R_y=1$ and $R_z=0.7$.
		}
		\label{figS2}
	\end{figure*}
	
	\begin{figure*}[t]
		\includegraphics[width=4in]{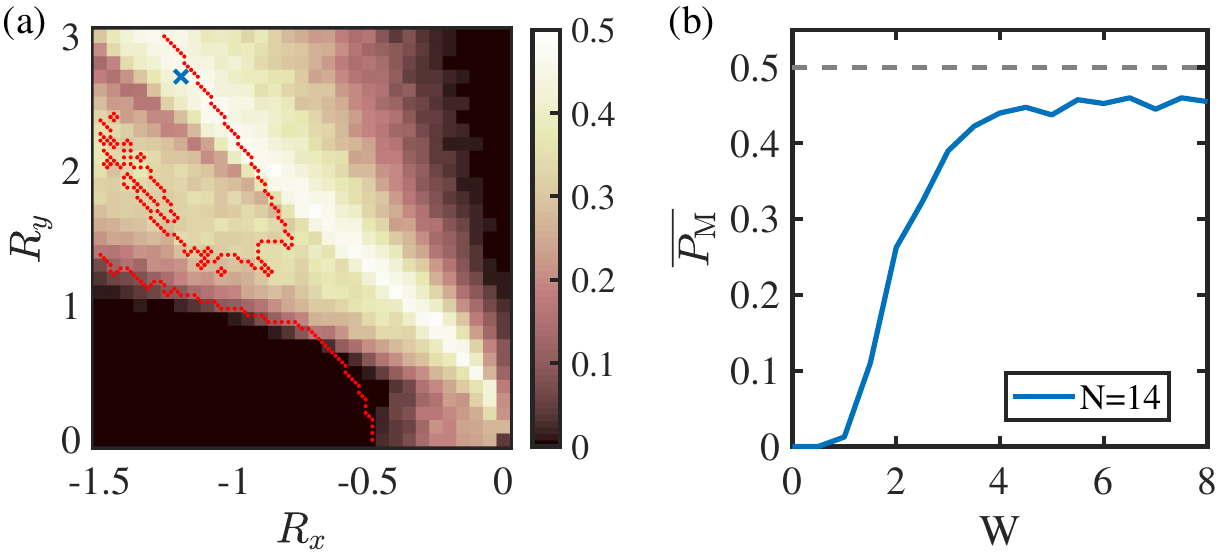}
		\caption{(Color online) 
			(a) $\mathbb{Z}_2$ invariant $\overline{P_\mathrm{M}}$ versus $R_x$ and $R_y$ for an amorphous lattice.
			For comparison, we also plot the phase boundaries for hard-core bosons at half-filling in a regular lattice as red dots.
			(b) $\overline{P_\mathrm{M}}$ versus disorder strength $W$ in a structurally disordered lattice when $R_x=-1.2$ and $R_y=2.7$ [marked in (a) by a blue diagonal cross].
			Here, $R_z=0.7$ and the system size $N=14$.
		}
		\label{figSS}
	\end{figure*}
	
	In this section, we plot $0.5-\overline{P_\mathrm{M}}$ as a function of the system size $N$ in the logarithmic scale 
	for $R_x=-0.25, -0.3, -0.325, -0.35$ in the topological regime for the many-body Hamiltonian [see Fig.~\ref{figS2}(a)]. 
	It shows an overall decrease of the value toward zero as the system size $N$ increases, suggesting that $\overline{P_\mathrm{M}}$ approaches $0.5$ in the thermodynamic limit.
	
	In Fig.~\ref{figS2}(b), we give the plot of an energy gap between the ground state and the first excited state 
	of the many-body Hamiltonian for different system sizes.
	We see that the energy gap decreases as the system size $N$ increases, suggesting that 
	the energy gap vanishes in the thermodynamic limit. 
	
	In Fig.~\ref{figSS}(a), we plot the phase diagram in the ($R_x$, $R_y$) plane for a larger system with $N=14$,
	illustrating the existence of a region where a topologically trivial phase in a regular lattice becomes nontrivial
	in an amorphous lattice. We further find that for a system parameter highlighted as a blue diagonal cross in Fig.~\ref{figSS}(a), $\overline{P_\mathrm{M}}$ undergoes a sharp change from $0$ to a value near $0.5$ as 
	the structural disorder strength $W$ increases [see Fig.~\ref{figSS}(b)], providing strong evidence of the existence of the structural disorder 
	induced topological phase transition in the many-particle level.
	
	\begin{table}[b]
		\caption{The system parameters used in the simulation for Fig.~2(e1-e4) and (f1-f4) in the main text. Here $d^2/a_0^3$ is the energy unit of the Rydberg Hamiltonian, $\Omega$ is the Rabi frequency of the microwave field, and $t_f$ is the evolution time.   }
		{\setlength\arrayrulewidth{0.5pt}
			\begin{tabular}{x{50pt}x{44pt}x{44pt}x{44pt}x{44pt}}
				\hline \hline
				Fig.~2:			& (e1,e2,f1)	& (f2) 		& (e3,e4,f3)	& (f4) 		\\ \hline
				$d^2/(a_0^3 h)$	& $10$ MHz		& $10$ MHz 	& $1$ MHz 		& $1$ MHz 	\\
				$\Omega/2\pi$	& $0.1$ MHz		& $0.3$ MHz	& $0.1$ MHz		& $0.2$ MHz	\\
				$t_f$			&$1.5~\mu$s		&$1.5~\mu$s	& $2~\mu$s		& $2~\mu$s	\\\hline\hline
		\end{tabular}}
		\label{tableS1}
	\end{table}
	
	\section{S-9. Numerical simulations for experimental observations of topological phases}
	To experimentally identify the topological phases, we apply a global microwave field with the Rabi frequency $\Omega(t)$ and detuning $\Delta(t)$ to couple the two Rydberg states, 
	which is described by the Hamiltonian,
	\begin{equation}
		\hat{\mathcal{H}}(t) = \hat{H} + \frac{\hbar \Omega(t)}{2}  \sum_{i=1}^{2N}(\hat{b}_i^\dagger + \hat{b}_i) - \hbar \Delta(t) \sum_{i=1}^{2N} \hat{b}_i^\dagger \hat{b}_i, 
	\end{equation} 
	where $\hat{H} = \sum_{i<j}^{2N} V_{ij} (\hat{b}_i^{\dagger} \hat{b}_j + \hat{b}_j^{\dagger} \hat{b}_i)$ is the Rydberg Hamiltonian defined in Eq.~(1) in the main text. 
	Starting from an empty state $\ket{0}$ with all atoms in the $s$-level, the state evolves as $\ket{\varphi(t)}=\hat{\mathcal{U}}(t,0) \ket{0}$ where $\hat{\mathcal{U}}(t,t_0)=\mathcal{T} [e^{-(i/\hbar) \int_{t_0}^{t} \hat{\mathcal{H}}(\tau) \mathrm{d}\tau}]$ with $\mathcal{T}[\cdots]$ being the time-ordering operator. 
	We use the Krylov subspace method to calculate the time evolution of the system.
	In the main text, we set the energy unit $d^2/a_0^3=1$ to simplify notations. Here in the simulation, 
	considering the realistic experimental parameters, we set $d^2/(a_0^3 h)$ to $1$ MHz or $10$ MHz.
	
	\subsection{A. The single-particle case}
	In the single-particle case, we shine a weak time-independent microwave radiation with the Rabi frequency $\Omega(t)=\Omega$ and detuning $\Delta(t)=\Delta$
	for several microseconds to excite an excitation. Table~\ref{tableS1} lists the energy unit $d^2/a_0^3$, the Rabi frequency $\Omega$ and the evolution time $t_f$
	used in the numerical simulation for Fig.~2(e1-e4) and (f1-f4) in the main text. Note that in the trivial cases in Fig.~2(f2) and (f4), we take a relatively 
	larger Rabi frequency in the time evolution in order to perform the postselection at $\Delta=0$; otherwise, there are only a few excitations,
	which can also be seen in Fig.~2(e2) and (e4) in the main text.
	
	\subsection{B. The many-body case}
	To measure the topological phases in the many-body case, we first prepare the system in the empty state $|0\rangle$ and then apply a global microwave pulse to couple the $|60S_{1/2}\rangle$ and $|60P_{1/2}\rangle$ levels with an initial microwave detuning $\Delta(t=0)=20$ MHz.
	In the simulation, we set $d^2/(a_0^3 h)$ to $10$ MHz, and $|0\rangle$ is approximately the highest energy many-body state of the Hamiltonian $\hat{H}-\hbar\Delta(t=0)\sum_{i=1}^{2N}\hat{b}_i^\dagger\hat{b}_i$. 
	We then slowly tune $\Omega$ and $\Delta$ according to the scheme shown in Fig. 3(c) to approximately drive the state to the highest energy state of the Hamiltonian $\hat{H}-\hbar\Delta_f\sum_{i=1}^{2N}\hat{b}_i^\dagger\hat{b}_i$, or the ground state of the Hamiltonian
	$-\hat{H}+\hbar\Delta_f\sum_{i=1}^{2N}\hat{b}_i^\dagger\hat{b}_i$. Hence, $-\Delta_f$ plays the role of the chemical potential for the Hamiltonian
	$-\hat{H}$. At the end, we measure the atom occupancy at edges or in the bulk on the $p$-level for the final state. 
	If $-\hat{H}$ is in the topological phase, then the edge sites occupancy should exhibit a sharp rise due to the emergence of particles mainly residing at
	the edges, when we vary $-\Delta_f$ across zero. This rise does not appear in the trivial phase. Here, the edge sites occupancy is defined as 
	the configuration averaged expectation value of $\hat{b}_1^\dagger\hat{b}_1$ or $\hat{b}_{2N}^\dagger\hat{b}_{2N}$,
	and the bulk sites occupancy is defined as the configuration averaged expectation value of $\sum_{j=2}^{2N-1} \hat{b}_j^\dagger\hat{b}_j/(2N-2)$. The characteristic signatures can therefore be utilized to diagnose 
	whether a system $-\hat{H}$ and thus $\hat{H}$ is in a topological phase. In fact, the ground state $\ket{\Psi_0}$ of  $\hat{H}$ at half filling is topological if and only if 
	the ground state $\ket{\Psi_0'}$ of $-\hat{H}$ is topological. 
	It is due to the fact that $\ket{\Psi_0'}=\hat{U}\ket{\Psi_0}$ because $\hat{U}\hat{H}\hat{U}^{\dagger}=-\hat{H}$ with $\hat{U}=\prod_{i=1}^{N} \sigma_{2i}^z$,
	leading to
	\begin{equation}
		\bra{\Psi_0'} \hat{\mathcal{P}}_\mathrm{M} \ket{\Psi_0'}= \bra{\Psi_0} \hat{U}^\dagger \hat{\mathcal{P}}_\mathrm{M} \hat{U} \ket{\Psi_0}
		=\bra{\Psi_0} \hat{\mathcal{P}}_\mathrm{M} \ket{\Psi_0}
	\end{equation}
	since $[\hat{U},\hat{\mathcal{P}}_\mathrm{M}]=0$.

\end{widetext}


\begin{thebibliography}{99}
\bibitem{footnote1}{This $\mathbb{Z}_2$ invariant is related to the polarization $P_\mathrm{S}$ for fermionic systems via the Jordan-Wigner transformation. 
The fact can be easily seen if we write the polarization $P_\mathrm{S}$ in a compact form as $P_\mathrm{S} = \frac{1}{2\pi} \mathrm{Im} \ln \bra{\Phi_0} \hat{\mathcal{P}}_\mathrm{S} \ket{\Phi_0}$ with $\hat{\mathcal{P}}_\mathrm{S} =  e^{\frac{2 \pi i}{N}  \sum_{j=1}^{2N} x_{j} (\hat{c}_{j}^\dagger \hat{c}_{j} -\frac{1}{2}) }$ and $\ket{\Phi_0}$ being the Slater determinant of all the occupied single-particle states.}
\end{thebibliography}
\end{document}